\documentclass[fleqn,10pt]{wlscirep}
\usepackage[utf8]{inputenc}
\usepackage[T1]{fontenc}
\usepackage{amsmath,amssymb}
\usepackage{bm}
\usepackage{graphicx}
\usepackage{color}

\usepackage{hyperref}
\hypersetup{
    colorlinks=true,       % false: boxed links; true: colored links
    linkcolor=blue,          % color of internal links
    citecolor=blue,        % color of links to bibliography
    filecolor=blue,      % color of file links
    urlcolor=blue           % color of external links
}

\title{Quantum annealing for systems of polynomial equations}

\author[1,*]{Chia~Cheng~Chang}
\author[2]{Arjun~Gambhir}
\author[3]{Travis~S.~Humble}
\author[4]{Shigetoshi~Sota}

\affil[1]{RIKEN Interdisciplinary Theoretical and Mathematical Sciences (iTHEMS), Wako, Saitama 351-0198, Japan}
\affil[1]{Department of Physics, University of California, Berkeley, California 94720, USA}
\affil[1]{Nuclear Science Division, Lawrence Berkeley National Laboratory, Berkeley, California 94720, USA}
\affil[2]{Nuclear and Chemical Sciences Division, Lawrence Livermore National Laboratory, Livermore, CA 94550, USA}
\affil[3]{Quantum Computing Institute, Oak Ridge National Laboratory, Oak Ridge, Tennessee 37831, USA}
\affil[4]{RIKEN Computational Materials Science Research Team, Kobe, Hyogo 650-0047, Japan}
\affil[*]{chiacheng.chang@riken.jp}

\begin{abstract}
Numerous scientific and engineering applications require numerically solving systems of equations. Classically solving a general set of polynomial equations requires iterative solvers, while linear equations may be solved either by direct matrix inversion or iteratively with judicious preconditioning. However, the convergence of iterative algorithms is highly variable and depends, in part, on the condition number. We present a direct method for solving general systems of polynomial equations based on quantum annealing, and we validate this method using a system of second-order polynomial equations solved on a commercially available quantum annealer. We then demonstrate applications for linear regression, and discuss in more detail the scaling behavior for general systems of linear equations with respect to problem size, condition number, and search precision. Finally, we define an iterative annealing process and demonstrate its efficacy in solving a linear system to a tolerance of $10^{-8}$.
\end{abstract}

\begin{document}

\flushbottom
\maketitle

\noindent\textbf{INTRODUCTION.}

Many problems in science, engineering, and mathematics can be reduced to solving systems of equations with notable examples in modeling and simulation of physical systems, and the verification and validation of engineering designs. Conventional methods for solving linear systems range from exact methods, such as matrix diagonalization, to iterative methods, such as fixed-point solvers, while polynomial systems are typically solved iteratively with homotopy continuation. The advent of quantum computing has opened up the possibility of new methods for solving these challenging problems. For example, a quantum algorithm for solving systems of linear equations was established for gate-based quantum computers~\cite{2009PhRvL.103o0502H} and demonstrated with small-scale problem instances \cite{zheng2017solving}. Additionally, an algorithm for solving linear systems within the adiabatic quantum computing model~\cite{subasi2018quantum} was experimentally demonstrated~\cite{wen2018experimental}, followed by a more recent proposal~\cite{2018Lomanaco}.
\par
In this work, we present an approach for solving a general system of $n^{\textrm{th}}$-order polynomial equations based on the principles of quantum annealing, followed by a demonstration of the algorithm for a system of second-order polynomial equations on commercially available quantum annealers. We then narrow the scope to examples of linear equations by first demonstrating an application to linear regression, before elucidating results on ill-conditioned linear systems motivated by the discretized Dirac equation $D\phi=\chi$ from lattice quantum chromodynamics (QCD). The solution to the discretized Dirac equation is currently the only approach for evaluating non-perturbative QCD. However, well-known numerical challenges slow convergence with conventional solvers~\cite{Stathopoulos:2007zi,AbdelRehim:2009by}. We end by using quantum annealing to solve a similar system and characterize the performance from experimental demonstrations with a commercial quantum annealer.\\

\noindent{\textit{\textbf{Polynomial Systems of Equations.}}}

We consider the system of $N$ polynomial equations
\begin{align}
F_i = P^{(0)}_i + \sum_j P^{(1)}_{ij} x_j + \sum_{jk} P^{(2)}_{ijk} x_j x_k + \dots = 0
\label{eq:system_poly}
\end{align}
where $i\in\{1,\dots, N\}$, and $P^{(n)}$ is a rank $n+1$ tensor of known real-valued coefficients for the polynomial of order $n$, and the real-valued vector $x$ denotes the solution.  Truncating to first order recovers a linear system of equations, i.e., $P^{(0)}_i+\sum_j P^{(1)}_{ij} x_j=0$. 
\par
Prior to this work, there exists no direct methods for solving a general $n^{\textrm{th}}$-order polynomial system. For linear systems, existing approaches include direct diagonalization using Gauss-Jordan elimination or iterative methods such as conjugate-gradient. In practice, direct diagonalization is limited in computational efficiency, as those methods scale sharply with the size of the matrix. By contrast, iterative methods may have greater computational efficiency but the performance and stability are often sensitive to the input matrix. 
Preconditioning improves convergence of linear systems by transforming the input as 
$M^{-1}P^{(1)}x=M^{-1}b$,
where the preconditioner $M$ must be inexpensive to invert and $M^{-1}$ should be ``close" to $P^{(1)^{-1}}$, so that $M^{-1}P^{(1)}$ resembles a matrix close to unity.
Identifying an effective preconditioner plays an important role in numerical convergence of iterative methods~\cite{1126-6708-2007-07-081,Brandt2008MultiLevelAS, Brannick:2007ue, 2013arXiv1303.1377F, Clark:2016rdz}. For lattice QCD applications~\cite{Chang:2018uxx}, the low-lying spectrum of the Dirac operator slows iterative convergence and preconditioning has been used to project out these low-lying modes. Acquiring the low-lying eigenpairs or singular triplets of $D$ is in general computationally expensive and requires the use of additional iterative methods that also suffer from critical slowing down. Solutions to address this issue include EigCG \cite{Stathopoulos:2007zi, AbdelRehim:2009by}, inexact deflation \cite{1126-6708-2007-07-081}, and adaptive multigrid \cite{Brandt2008MultiLevelAS, Brannick:2007ue, 2013arXiv1303.1377F, Clark:2016rdz}.\\

\noindent{\textbf{RESULTS}}

\noindent\textit{\textbf{Quantum Annealing for Polynomial Solvers}}

Quantum annealing offers an alternative approach to solving a general system of equations. We map each variable $x_{j}$ using $R$ number of qubits such that
\begin{equation}
x_j = a_j \sum_{r=0}^{R-1} 2^r \psi_{rj} + b_j.
\label{eq:xmapping}
\end{equation}
where $\psi_{rj} \in \{0,1\}$, $a_i \in \mathbb{R}$ and $b_i \in \mathbb{R}$ such that $x_{j} \in \{b_j+2^{r-1}a_j | r \in \mathbb{Z}^\geq < R\}$. Defining the vectors
\begin{align}
\mathcal{A}&\equiv \begin{pmatrix} a_0 & \dots & a_{N-1}\end{pmatrix} && |\mathcal{A}| = N \nonumber \\
\mathcal{B}&\equiv \begin{pmatrix} b_0 & \dots & b_{N-1}\end{pmatrix} && |\mathcal{B}| = N \nonumber \\
\mathcal{R}&\equiv \begin{pmatrix} 2^0 & \dots & 2^{R-1}\end{pmatrix} && |\mathcal{R}| = R \nonumber \\
\mathcal{X}&\equiv \begin{pmatrix} x_0 & \dots & x_{N-1} \end{pmatrix} && |\mathcal{X}| = N \nonumber\\
\Psi  &\equiv \begin{pmatrix} \psi_{00} & \dots & \psi_{R-1\ N-1} \end{pmatrix} && |\Psi| = N\times R \nonumber 
\end{align}
where $|V|$ is the cardinality operator yielding the number of elements in a generic vector $V$. The objective function $\chi^2$ which solves Eq.~(\ref{eq:system_poly}) is given by minimizing the residual sum of squares in the qubit-basis
\begin{align}
\chi^2 = &\left[P^{(0)} + P^{(1)}(\cdot \mathcal{B} + \circ \mathcal{A} \otimes \mathcal{R} \cdot \psi) + P^{(2)}(\cdot \mathcal{B} + \circ \mathcal{A} \otimes \mathcal{R} \cdot \psi)^2 + \dots\right]^2,\label{eq:system_poly_optimize} \\
\equiv & Q^{(0)} + Q^{(1)} + \dots + Q^{(2N)} \label{eq:qn_definition}
\end{align}
where $\cdot$ is the dot product, $\circ$ is the Hadamard product, and $\otimes$ is the tensor product. In particular, $(\circ \mathcal{A} \otimes \mathcal{R})^n \equiv \circ \mathcal{A}^{\otimes n} \otimes \mathcal{R}^{\otimes n}$, where $V^{\otimes n}$ is a repeated $n$ sequence of tensor products. The ground state of Eq.~(\ref{eq:system_poly_optimize}) solves a system of polynomial equations. For current commercial quantum annealers, auxiliary qubits are required to reduce multi-linear terms down to bilinear interactions through quadratization~\cite{Rosenberg1975, BOROS2002155, PhysRevA.77.052331, Boros2012, humble2014integrated, 2015arXiv150804816T, 2015arXiv150807190O, 2019arXiv190104405D}. We provide the details of quadratization through reduction-by-substitution on a system of second-order polynomials in Methods.

Finally, we note that resulting energy at the end of the optimization corresponds to exactly the residual sum of squares if the constant terms in $\chi^2$ are correctly accounted for. It follows that the entire energy spectrum is positive, and if the exact solution is recovered, then the ground state energy must be zero.\\

\noindent\textit{\textbf{Quantum Annealing for Linear Solvers}}

A system of linear equations simplifies Eq.~(\ref{eq:system_poly_optimize}) to involve only bilinear terms without quadratization, and reduces to a quadratic unconstrained binary optimization (QUBO) problem where $H^{\mathrm{QUBO}}(\psi) = \sum_{ij}\psi_i Q_{ij} \psi_j$ with
\begin{align}
Q =  &
\begin{pmatrix}
a_1^2 P^{(1)}_{11}
& \dots &
a_1 a_{N} P^{(1)}_{1N}\\
\vdots & \ddots & \vdots \\
a_{N} a_1 P^{(1)}_{N1} & \dots & a_{N}^2 P^{(1)}_{NN}
\end{pmatrix}
\otimes 
\begin{pmatrix}
2^0 2^0 & \dots & 2^0 2^{R-1} \\
\vdots & \ddots & \vdots\\
2^{R-1} 2^0 & \dots & 2^{R-1} 2^{R-1}
\end{pmatrix}
+ 2 \begin{pmatrix}
a_1 P^\prime_{1} \\
& \ddots & \\
& & a_{N} P^\prime_{N}\\
\end{pmatrix}
\otimes
\begin{pmatrix}
2^0 & & \\
& \ddots & \\
& & 2^{R-1}
\end{pmatrix}
\label{eq:qubo_hamiltonian}
\end{align}
where $P^\prime_{n} = P^{(0)}_{n} + \sum_i b_i P^{(1)}_{ni}$.  In addition, constant terms that arise from the substitution of Eq.~(\ref{eq:xmapping}) are omitted for simplicity and leaves the solution vector $\Psi$ unchanged, but should be included when interpreting the energy as the residual sum of squares.\\

\noindent\textit{\textbf{Application to Linear Regression}}

Given a set of $N$ identical and independent observations of the
\begin{align}
\mathrm{independent\ } \{x_i : i\in\{1,...,X\}\} && \mathrm{dependent\ } \{y_{i;g} : i\in\{1,...,X\}, g\in\{1,...,N\}\} \nonumber
\end{align}
variable, the mean and covariance of $y_i$ follows
\begin{align}
\langle y_i \rangle = \frac{1}{N} \sum_{g=1}^{N} y_{i;g} && S_{ij} = \langle(y_i - \langle y_i \rangle)(y_j - \langle y_j \rangle) \rangle  \nonumber
\end{align}
where the angle brackets denote the expectation value over $N$ observations. A fitting function $F(x_i,p)$ may be defined with respect to the set of $P$ unknown parameters $ p = \{p_n : n\in\{1,...,P\}\} $, and a corresponding objective function for generalized least squares may be defined as
\begin{align}
\sum_{ij}\left[F(x,p) - \langle y \rangle\right]_i S^{-1}_{ij} \left[F(x,p) - \langle y \rangle\right]_j
\label{eq:chi2}
\end{align}
where the optimal value for the set $p$ is determined by minimizing Eq.~(\ref{eq:chi2}).
\par
Restriction to linear least squares demands that the fitting function is linear in the unknown parameters, and therefore may be written in the form
\begin{equation}
F(x_i,p) = \sum_{n=1}^{P} p_n f_n(x_i)
\label{eq:reg_fcn}
\end{equation}
where $f_n(x_i)$ can be any function. The solution for linear regression is obtained by expanding Eq.~(\ref{eq:chi2}) with Eq.~(\ref{eq:reg_fcn}) and yields
\begin{align}
\sum_{ij}\left[ \sum_n p_n f_n(x_i) - y_i \right] S_{ij}^{-1} \left[\sum_m p_m f_m(x_j) -y_j\right].
\label{eq:chi2_exp}
\end{align}
The extrema of the objective function can be determined by taking the derivative of Eq.~(\ref{eq:chi2_exp}) with respect to $p_n$ yielding a matrix equation of the form $\sum_j A^{(1)}_{ij} p_j = A^{(0)}_i$ analogous to Eq.~(\ref{eq:system_poly}) where
\begin{align}
P^{(1)} = 
\begin{pmatrix}
f_0(x)^T S^{-1} f_0(x) & \dots & f_0(x)^T S^{-1} f_{P}(x)\\
\vdots & \ddots & \vdots\\
f_{P}(x)^T S^{-1} f_0(x) & \dots & f_{P}(x)^T S^{-1} f_{P}(x)
\end{pmatrix} &&P^{(0)} = 
\begin{pmatrix}
f_0(x)^TS^{-1}y\\
\vdots\\
f_{P}(x)^T S^{-1}y\\
\end{pmatrix}.
\label{eq:classical_direct}
\end{align}
The solution to least squares minimization can then be mapped to a QUBO problem following Eq.~(\ref{eq:qubo_hamiltonian}), and amenable to methods of quantum annealing.\\

\noindent{\textbf{DISCUSSION}}

\noindent\textbf{\textit{System of Second Order Polynomial Equations}}

We demonstrate the validity of the algorithm on a system of two second order polynomial equations. The problem is chosen to be small such that the solution can be confirmed by an explicit search over the entire Hilbert space, and evaluated onto a D-Wave annealer. Consider the following system of equations,
\begin{align}
0 = 2x_0^2 + 3x_0 x_1 + x_1^2 + 2x_0 + 4x_1 - 51, && 0 =  x_0^2 + 2x_0x_1 + 2x_1^2 + 3x_0 + 2x_1 - 46, \nonumber
\end{align}
with four real solutions at
\begin{align}
(x_0, x_1) = (2, 3), && (x_0, x_1) \approx (-10.42, 7.27), && (x_0, x_1)  \approx (-3.29, -3.74), && (x_0, x_1) \approx & (7.71, -3.53).\nonumber
\end{align}
For the sake of discussion, we set up Eq.~(\ref{eq:system_poly_optimize}) to solve for the solution at $(2, 3)$ by choosing $\mathcal{A} = \begin{pmatrix} 1 & 1 \end{pmatrix}$, $\mathcal{B} = \begin{pmatrix} 0 & 0 \end{pmatrix}$, and $\mathcal{R} = \begin{pmatrix} 2^0, 2^1\end{pmatrix}$. The tensors $P^{(n)}$ are obtained by inspection,
\begin{align}
P^{(0)} = \begin{pmatrix} -51 \\ -46 \end{pmatrix}, &&P^{(1)} = \begin{pmatrix}2 & 4 \\ 3 & 2 \end{pmatrix},
&& P^{(2)} = \begin{pmatrix}\begin{pmatrix} 2 & 3 \\ 0 & 1 \end{pmatrix} & \begin{pmatrix}1 & 2 \\ 0 & 2 \end{pmatrix}\end{pmatrix}. \nonumber
\end{align}
After transforming to the qubit-basis, direct search of the ground state of the 4-body Hamiltonian yields
\begin{equation}
\Psi = \begin{pmatrix} 0 & 1 & 1 & 1 \end{pmatrix} \rightarrow \mathcal{X} = \begin{pmatrix} 2 & 3\end{pmatrix}\nonumber
\end{equation}
where the consecutive pairs of binary variables maps to the binary representation of $x_i$ with little-endianness due to specific choices of $\mathcal{A}$, $\mathcal{B}$ and $\mathcal{R}$. The solution is reproduced when transforming the set of generalized $n$-dimensional $\mathcal{Q}\equiv \{Q^{(0)}, Q^{(1)}, \dots, Q^{(N)}\}$ matrix as defined in Eq.~(\ref{eq:qn_definition}), to upper-triangular tensors, and also reproduced when further reducing the dimensionality of elements in $Q^{(n\geq 2)}$ with repeated indicies, yielding in general the most sparse upper-triangular representation $\mathcal{Q}^{\textrm{sparse}}$.

We quadratize the $\mathcal{Q}^{\textrm{sparse}}$ set of rank 0 to $N$ tensors to the QUBO representation with reduction-by-substitution ~\cite{Rosenberg1975, BOROS2002155, PhysRevA.77.052331, 2019arXiv190104405D} by introducing $\frac{1}{2}N(N-1)$ auxiliary qubits $\psi^a$ to enforce the following constraint,
\begin{equation}
C(\psi_i \psi_j -2\psi_i \psi^a_{ij} - 2\psi_j \psi^a_{ij} + 3 \psi^a_{ij})
\label{eq:constraints}
\end{equation}
such that the constraint is minimized when $\psi^a_{ij} = \psi_i \psi_j$. The coefficient $C$ should be chosen large enough such that the constraint is satisfied under optimization. Additional details of the quadratization used is given in Methods.

We repeat the exercise of solving the same system of polynomial equations on the D-Wave annealer with the symmetrized and quadratized representation, and successfully reproduce the solution,
\begin{align}
\Psi = & \begin{pmatrix}\psi_0 & \psi_1 & \psi_2 & \psi_3 & \psi^a_{01} & \psi^a_{02} & \psi^a_{03} & \psi^a_{12} & \psi^a_{13} & \psi^a_{23}\end{pmatrix}\nonumber \\
= & \begin{pmatrix}0 & 1 & 1 & 1 & 0 & 0 & 0 & 1 & 1 & 1 \end{pmatrix} \rightarrow \mathcal{X} = \begin{pmatrix} 2 & 3 \end{pmatrix}\nonumber
\end{align}
where the last six auxiliary qubits confirm consistency of reducing many body interactions down to bilinear terms. Finally, we mention that with reduction-by-substitution, a system of $m^{\mathrm{th}}$-order polynomials needs to be quadratized $m-1$ times, requiring exponentially more auxiliary qubits. 

The software implementation of various many-body and 2-body quadratized representations for a system of second-order polynomial equations, the brute force solver, and D-Wave solver are made publicly available~\cite{Chang19github}.\\

\noindent\textbf{\textit{Linear Regression}}

As an example, consider the following artificially generated data
\begin{align}
\mathrm{E}\left[D(x)\right]  = 8 + 4x + 7x^2 \label{eq:lls_mean} && \mathrm{Var}\left[D(x)\right] = \mathrm{E}\left[D(x)\right]/10 && \mathrm{Corr}\left[D(x_i),D(x_j)\right] = & 0.9^{\left|x_i-x_j\right|}
\end{align}
where $x \in \Bbb Z : x\in [0,49]$. The Toeplitz correlation matrix~\cite{2011arXiv1106.5834H} is chosen to simulate a correlated time-series dataset, where the correlations decay exponentially as a function of $x$. Following the notation in Eq.~(\ref{eq:reg_fcn}), we assume a linear fit 
\begin{equation}
F(x, A) = A_0 + A_1 x + A_2 x^2,
\end{equation}
and we estimate the parameters $A_{n}$ given the data $D(x)$. Using Eq.~(\ref{eq:xmapping}), we express each parameter $A_i$ as a 4-bit unsigned integer
\begin{equation}
A_i = \psi_{1i} + 2\psi_{2i} + 4\psi_{3i} + 8\psi_{4i}
\end{equation}
and construct the problem Hamiltonian following Eqs.~(\ref{eq:qubo_hamiltonian}) and (\ref{eq:classical_direct}). The required 12 logical qubits (3 parameters $\times$ 4-bit representation) support a total of 4096 possible solutions. Explicit evaluation finds the true ground-state to have energy $E_0 = -1.418$ and eigenstate 
\begin{align}
\Psi_0 =& \left( 0\ 0\ 0\ 1\ 0\ 0\ 1\ 0\ 1\ 1\ 1\ 0 \right)
\label{eq:lls_gs}
\end{align}
which corresponds to the parameter values 
\begin{align}
A_0 = \left(0\ 0\ 0\ 1 \right) \rightarrow 8 && A_1 =  \left(0\ 0\ 1\ 0 \right) \rightarrow 4 && A_2 =  \left(1\ 1\ 1\ 0 \right) \rightarrow 7\nonumber
\end{align}
These correct coefficients for the generating function in Eq.~(\ref{eq:lls_mean}) verify the design of the algorithm.
\par
We next test the algorithm by solving the objective function using quantum annealing. The target Hamiltonian of Eq.~(\ref{eq:lls_mean}) is solve with a D-Wave annealer, and results for 100,000 independent evaluations are acquired using an annealing schedule with $T = 200~\mu$s. The correct result is reproduced in 0.5\% of the solves, while the lowest 0.8\% of the eigenvalue spectrum is obtained by 10\% of the solves with overall results biased towards the lower-lying eigenspectrum.\\

\noindent\textit{\textbf{Conditioned Systems of Linear Equations}}

\label{sec:gle}
In this section we show results and scaling of a classical method and the quantum annealer. One of the criteria for categorizing the ``difficulty" of a linear system is condition number.  The condition number of a matrix $P^{(1)}$ is defined as the ratio of maximum and minimum singular values. 
\begin{equation}
\kappa(P^{(1)})=\frac{\sigma_{\operatorname{max}}(P^{(1)})}{\sigma_{\operatorname{min}}(P^{(1)})}
\end{equation}
In the case of symmetric matrices, this is equivalent to the ratio of largest and smallest eigenvalues.

We vary our test matrices in two ways: 1) vary the problem size while holding the condition number fixed, 2) the problem size is held constant with varying condition number. The accurately of the solution will be judged by the relative residual sum of squares,
\begin{equation}
\chi^2_{\textrm{rel.}} = \frac{(P^{(1)}x_{\operatorname{approx}}+P^{(0)})^2}{P^{(0)^2}} = \frac{E_0}{P^{(0)^2}},
\label{relativeresid_eq}
\end{equation}
where $E_0$ is the ground-state energy. For conjugate gradient, a tolerance for Eq.~(\ref{relativeresid_eq}) is utilized as a terminating criterion and the number of iterations when this point is reached is recorded. For quantum annealing the role of the relative residual is more subtle. The annealer is run many times and the lowest energy eigenpair is returned. The eigenvector from this set is substituted for $x_{\operatorname{approx}}$, allowing a relative residual to be defined for the total anneal.\\

\noindent\textit{{Classical Solutions}}

For the examples with a classical linear solver, conjugate gradient is used on $N = 12$, with varying condition number. Although conjugate gradient is not the optimal choice for classically solving such systems, the scaling comparison in condition number with the quantum algorithm is informative. Fig.~\ref{fig:iter_vs_cond} shows slightly worse than square root scaling of conjugate gradient with condition number.

The matrices from this and subsequent results are constructed by creating a random unitary matrix of rank $N$, denoted as $U$. A diagonal matrix $\Lambda$ is then linear populated by evenly spaced real-valued eigenvalues, such that $\mathrm{max}(\Lambda)/\mathrm{min}(\Lambda)=\kappa$. The matrices are trivially formed as $P^{(1)}=U\Lambda U^\dagger$. A common right-hand side is taken for all $P^{(1)}$: a vector of length $N$ with linearly spaced decimals between 1 and -1.\\

\noindent\textit{Quantum annealing}

In the following section, we demonstrate the scaling of the annealing algorithm under varying problem size, condition number, and precision of the search space. We conclude by applying the algorithm iteratively on a fixed problem and study the convergence of the relative residual.

\textit{Problem size --} 
In Fig.~\ref{fig:axb_size}a, we study the scaling behavior for $\kappa = 1.1$ and $R=2$. Due to prior knowledge of the conjugate-gradient solution, the search space for all $N$ parameters are fixed for the set of problems, and encompass the minimum and maximum results of the solution vector $x$. Additionally, knowledge of the result allows us to identify the ground-state QUBO solution by minimizing the difference between the conjugate-gradient and QUBO results (the forward error), and studies the theoretical scaling of the algorithm absent of current hardware limitations. Due to the small condition number of this study, minimizing the forward error is equivalent to minimizing the backwards error.

With increased problem size, we observe that the percentage of annealed solutions which return the ground state decreases exponentially. This indicates the solution for a dense matrix may require exponentially more evaluations to obtain for current quantum annealers. The observed scaling is consistent with the assumption that the energy gap exponentially vanishes with increasing size for a dense Hamiltonian. In particular, beyond $n=16$, only one out of 100,000 evaluations yield the resulting annealed solution, demonstrating that the real ground-state is well beyond the reach of the available statistics.

\textit{Condition number --} Fig.~\ref{fig:axb_size}b demonstrates the scaling of the algorithm with respect to changing condition number. The problem size is fixed to $N=12$, and $R=2$. The condition number affects the solution vector $x$, and therefore for this study we restrict the search range to span exactly the minimum and maximum values of $x$. The chosen search range keeps the resulting relative residual approximately constant under varying condition number. For linear systems with larger condition numbers, minimizing the forward error is no longer a reliable estimate of the residual of the backwards error, and is therefore dropped from this study.

With increasing condition number, we observe that the percentage of solutions that converge to the lowest-lying state is a relatively constant value as demonstrated by a less than one order-of-magnitude change between the different examples. This behavior is in stark contrast with the scaling observed in Fig.~\ref{fig:axb_size}a, and suggests that with increasing condition number, the ground state is exponentially easier to identify. This is in amazing contrast to the classical result from Fig.~\ref{fig:iter_vs_cond}, in which convergence to the solution decreases as condition number is raised.

\textit{Precision of search --} Fig.~\ref{fig:axb_size}c explores the behavior of the algorithm as $R$ is increased for $N=4$ and $\kappa=1.1$. We observe that the relative residual exponentially decreases, as expected due to sampling an exponential number of solutions. However, increasing $R$ also requires exponentially more evaluations from the annealer in order to resolve the ground state. Similarly to Fig.~\ref{fig:axb_size}a, we observe that the forward error for problem sizes beyond $R=5$ starts to deviate from the backward error, an indication that the limits of hardware control have been reached.

\textit{Iterative approach --} Finally, we explore the possibility of iteratively applying the algorithm in order to decrease the relative residual of the final solution. We demonstrate this technique on $N=4$, with $\kappa=1.1$, and $R=4$. For this study, we initially set $\textrm{min}(\mathcal{X}) = -1$ and $\textrm{max}(\mathcal{X}) = 1$. With each iteration, we narrow the optimization to two neighboring values of the result allowed by the search space. Fig.~\ref{fig:axb_search} shows how the search space is refined with each iteration of the algorithm and converges to the conjugate gradient solution. Fig.~\ref{fig:axb_size}d shows that the relative residual exponentially decreases with the application of each iteration, while the number of anneals required to sample the ground state stays relatively constant. The solution from quantum annealing at the final (ninth) iteration agrees with conjugate-gradient at single precision accuracy.\\

\noindent\textbf{METHODS}

\noindent{\textit{\textbf{Quadratization}}}

Quadratization (\textit{i.e.} to make quadratic) maps terms in the Hamiltonian that are multi-linear with respect to the binary variables $\mathcal{X}$, to a larger Hilbert space involving only bilinear (\textit{i.e.} quadratic) contributions. This transformation is required to realize quantum algorithms with non-linear operations on near-term quantum computers. There exists in current literature, a rich selection of methods to perform such a task~\cite{Rosenberg1975, BOROS2002155, PhysRevA.77.052331, Boros2012, humble2014integrated, 2015arXiv150804816T, 2015arXiv150807190O, 2019arXiv190104405D}. For this work, we apply reduction-by-substitution~\cite{Rosenberg1975, BOROS2002155,2019arXiv190104405D}, where the constraint equation is given by Eq.~(\ref{eq:constraints}).

One constraint equation is required to define each auxiliary qubit $\psi_{ij}$. After taking into consideration that in the qubit basis, the Hamiltonian is symmetric under permutations of all indices, $\frac{1}{2}N^\prime(N^\prime-1)$ auxiliary qubits are needed to account for every unique quadratic combination of the underlying basis of length $N^\prime$.  In Fig.~(\ref{fig:egqubo}), we provide the smallest non-trivial example which maps a system of two second-order polynomial equations (\textit{i.e.} $N=2$), with $R=2$ after the $n$-body Hamiltonian has been reduced to the set $\mathcal{Q}^{\textrm{sparse}}$. The constraint equations have an overall coefficient $C$ which needs to be large enough such that the constraints are satisfied under optimization. For visual clarity, the constraints in the second quadrant are entered in the lower triangular section, but in practice should be accumulated with the upper triangular section occupied by $Q^{(2)}$.

Generalization to a system of more equation ($N>2$), or finer searches ($R>2$) is straightforward, and implemented in the accompanying software~\cite{Chang19github}. Generalizing to higher-order polynomial equations required additional levels of quadratization that was not implemented in this study.\\

\noindent{\textit{\textbf{Quantum Annealing}}}

Akin to adiabatic quantum optimization \cite{2000quant.ph..1106F}, quantum annealing prepares a quantum statistical distribution that approximates the solution by applying a slowly changing, time-dependent Hamiltonian~\cite{PhysRevB.39.11828, 1998PhRvE..58.5355K, RevModPhys.80.1061}, where measurements drawn from the distribution represent candidate solutions. Unlike adiabatic quantum computing, quantum annealing permits non-adiabatic dynamics at non-zero temperature, making this approach easier to realize experimentally but also more challenging to distinguish quantum mechanically~\cite{1998PhRvE..58.5355K, 2001Sci...292..472F, Dickson13, 2009PhRvA..79b2107A, 2015PhRvA..91f2320A, 2016PhRvB..93v4414W, 2016PhRvE..94c2105N, 2017PhRvP...8f4025M}. While examples of non-trivial advantages have been observed for fixed-size problem instances~\cite{ronnow2014defining, katzgraber2014glassy, denchev2016computational, PhysRevA.94.022337,PhysRevX.8.031016}, more general statements about computational complexity remain unresolved~\cite{mandra2018}. 

We demonstrate the proposed algorithm using the D-Wave 2000Q commercial quantum annealer. This hardware is based on cryogenically cooled superconducting electronic elements that implement a programmable Ising model. Each quantum register element expresses a single Ising spin variable, but the D-Wave 2000Q supports only a limited connectivity between these elements. In particular, the $i$-th spin variable may be assigned a bias $Q_{ii}$ and can be coupled to a unique set of six neighboring registers through the coupling $Q_{ij}$.  A densely connected Hamiltonian can be embedded into the hardware by using secondary constraints to build chains of strongly correlated elements in which $Q_{ij}^{\mathrm{constraint}} \gg Q_{ij}^{\mathrm{problem}}$. This coupling constraint favors chains of spin elements which behave as a single spin variable~\cite{2008arXiv0804.4884C}. Previous studies have identified optimal mappings of the infinite dimensional to three-dimensional Ising model~\cite{Choi2008,Klymko2014}. For the D-Wave 2000Q, approximately 64 logical spin variables may be represented within the 2048 physical spin elements. Our examples use the \texttt{dwave-sapi2} Python library~\cite{sapi2}, which is a software tool kit that facilitates cloud access to the annealer and supports a heuristic embedding method for the available hardware.\\

\noindent{\textbf{DATA AVAILABILITY}
	
Software to reproduce the solutions of the second-order system of polynomial equations discussed in this work is made publicly available at \href{https://github.com/cchang5/quantum_poly_solver}{https://github.com/cchang5/quantum\_poly\_solver}.\\

\noindent{\textbf{ACKNOWLEDGMENTS}

We thank Hidetoshi Nishimori (Tokyo Tech), Tetsuo Hatsuda (RIKEN), and Chih-Chieh Chen (ITRI) for discussions. This work was performed under the auspices of the U.S. Department of Energy by LLNL under Contract No. DE-AC52-07NA27344 (AG). Access to the D-Wave 2000Q computing system was provided by Oak Ridge National Laboratory. TSH acknowledges support from the Department of Energy, Office of Science, Early Career Research Project and the ORNL Directed Research and Development funds. This manuscript has been authored by UT-Battelle, LLC, under Contract No.~DE-AC0500OR22725 with the U.S.~Department of Energy. The Department of Energy will provide public access to these results of federally sponsored research in accordance with the DOE Public Access Plan (\href{http://energy.gov/downloads/doe-public-access-plan}{http://energy.gov/downloads/doe-public-access-plan}).\\

\noindent{\textbf{AUTHOR CONTRIBUTIONS}}

Initial idea was proposed by C.C.C.. Design of test cases was done by A.G. and C.C.C.. Conjugate-gradient solves was done by A.G.. Calculation done on the DWave 2000Q was performed by C.C.C. and T.S.H.. Cross checks with density matrix renormalization group was performed by S.S. and C.C.C.. All authors contributed to writing and editing of the final manuscript.\\

\noindent{\textbf{ADDITIONAL INFORMATION}}

\textbf{Competing Interests:} The authors declare no competing interests.

\bibliography{QA_LLS}

\begin{figure}[th]
	\centering
	\includegraphics[width=0.49\textwidth]{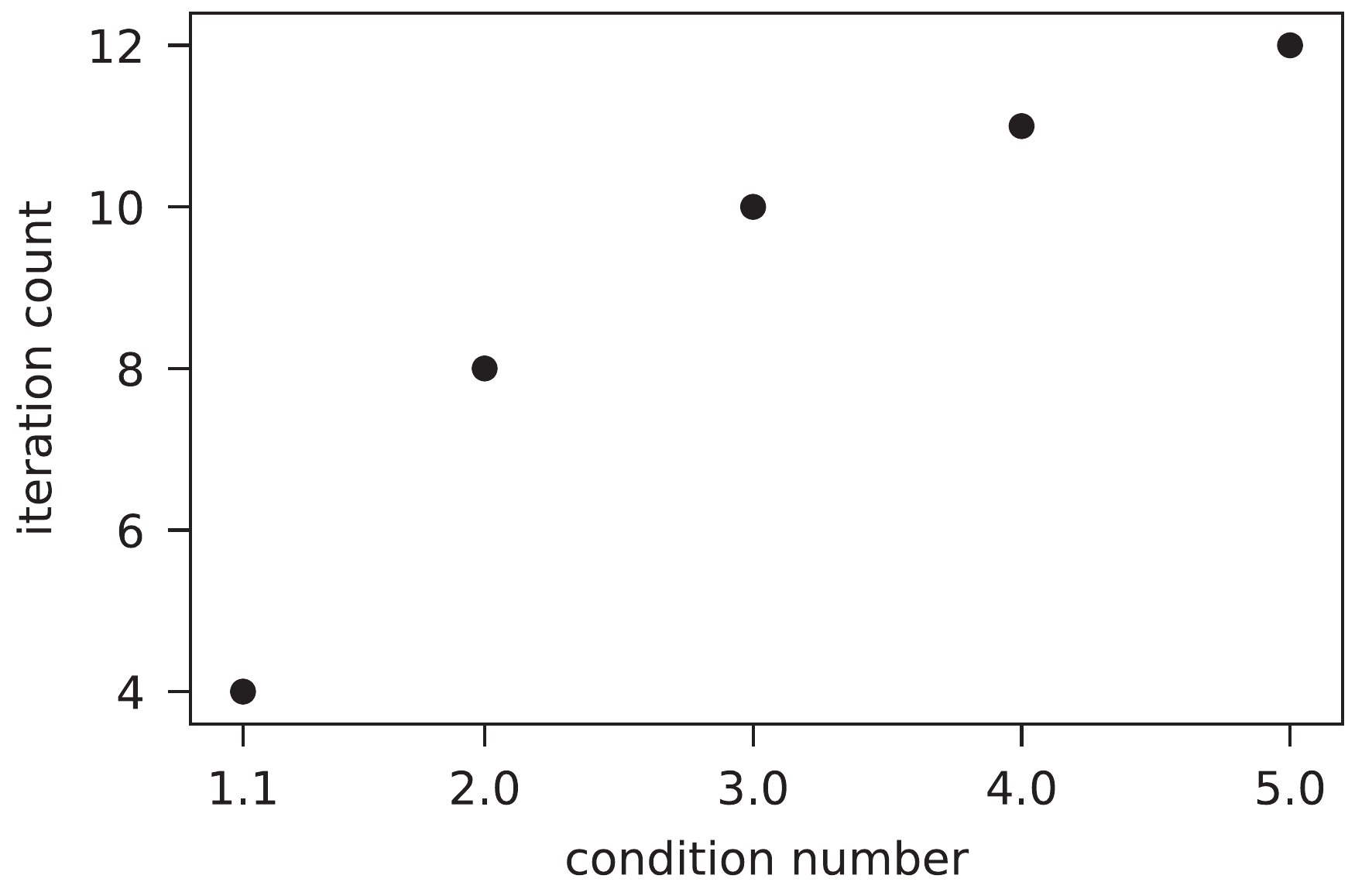}
	\caption{The number of conjugate gradient iterations grows slightly worse than $\sqrt{\kappa(P^{(1)})}$. The stopping criterion is a tolerance of $10^{-6}$ for the norm of the relative residual. All matrices are rank 12, with smaller eigenvalues as $\kappa(P^{(1)})$ increases, but identical eigenvectors. The same right-hand side is solved for all cases.}
	\label{fig:iter_vs_cond}
\end{figure}

\begin{figure*}[h]
	\includegraphics[width=\textwidth]{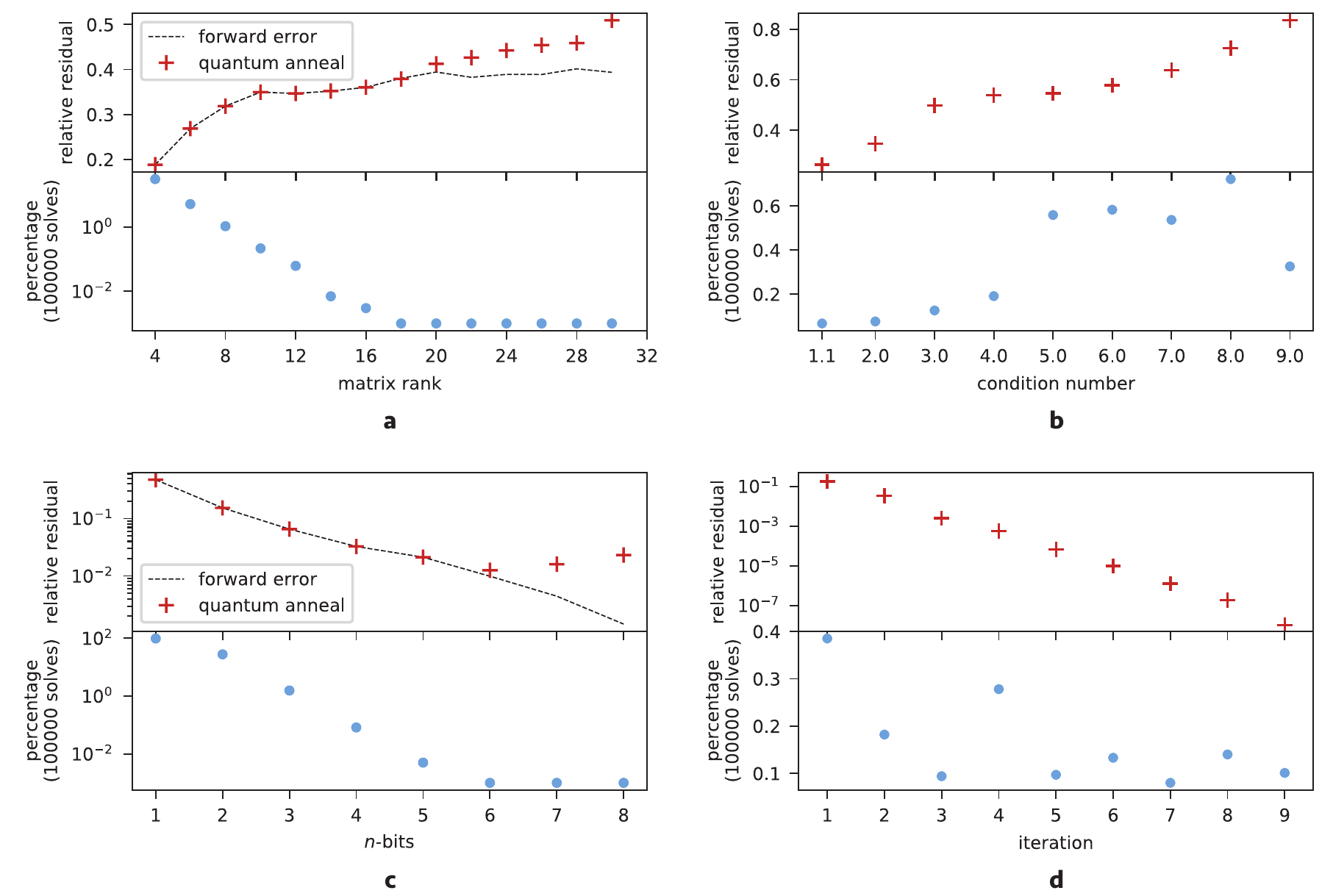}
	\caption{ \textbf{a,} (Top) The black dashed line shows the theoretical minimum relative residual of the algorithm as predicted by minimizing the forward error, given the search precision and search range used for the test. The red crosses are results of the lowest energy state from 100,000 quantum annealing measurements. Physical measurements deviate from the theoretical minimum as problem size grows.  (Bottom) The corresponding percentage of measurements in the minimum energy state are shown in blue. The vertical axis is shown in a logarithmic scale. \textbf{b,} Analogous to Fig.~\ref{fig:axb_size}a but for varying condition number. The forward error prediction (dashed black line) is omitted and not a reliable measure of the relative residual for larger condition numbers. Note that the vertical axis of the bottom plot showing the percentage of measurements observed in the lowest-lying state is on a linear scale. \textbf{c,} This plot is analogous to Fig.~\ref{fig:axb_size}a but for varying search precision. Note that both vertical axes are on a logarithmic scale. \textbf{d,} (Top) The relative residual exponentially decreases with each iteration of the algorithm. By the ninth iteration the result reaches single precision. (Bottom) The percentage of quantum annealing solutions in the lowest-lying state. The algorithm successfully resolves the solution at single precision for this example without issue.
	}
	\label{fig:axb_size}
\end{figure*}

\begin{figure*}[t]
	\includegraphics[width=\textwidth]{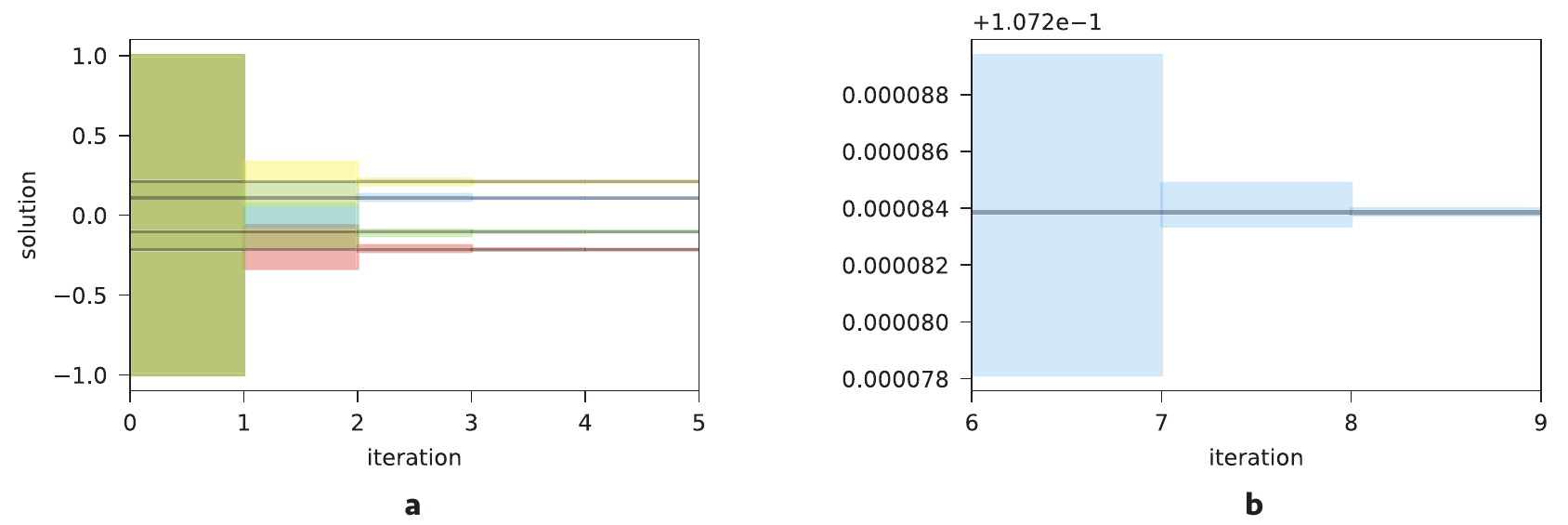}
	\caption{Search region over successive iterations of the algorithm. The shaded regions indicate the search region for each component of the solution vector $\mathcal{X}$. The gray horizontal line is the result from a conjugate-gradient solver. \textbf{a,} Search region for the first five iterations for all parameters. \textbf{b,} Search region zoomed in on a single parameter reaching single precision accuracy.
	}
	\label{fig:axb_search}
\end{figure*}

\begin{figure*}
	\centering
	\includegraphics[width=0.5\textwidth]{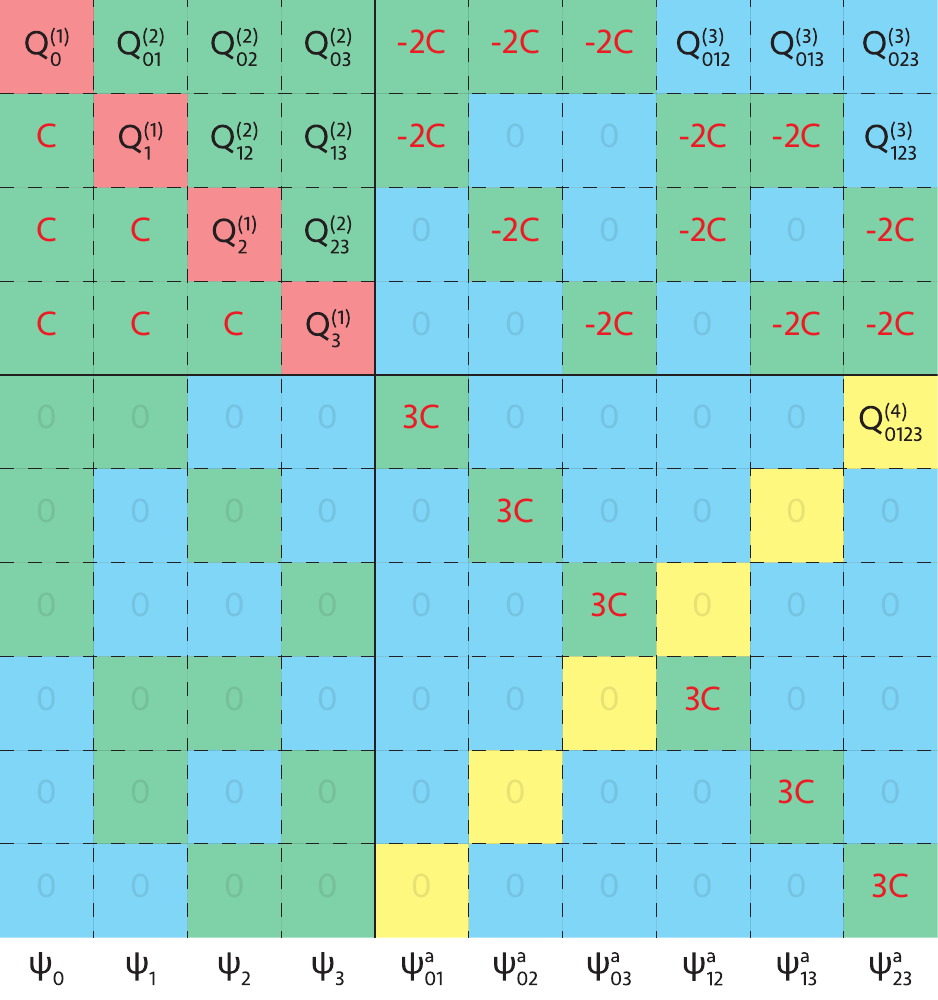}
\caption{Example QUBO for a system of second-order polynomial equations. The QUBO can be organized into four quadrants as indicated by the solid black lines, corresponding to bilinear (2${}^{\mathrm{nd}}$ quadrant), tri-linear (1${}^{\mathrm{st}}$ and 3${}^{\mathrm{rd}}$ quadrants) and quadra-linear (4${}^{\mathrm{th}}$ quadrant) contributions. Within the quadrants, the elements are colored to reflect the effective one (red), two (green), three (blue), and four (yellow) qubit interactions after accounting for repeated indices. The entries in $\mathcal{Q}^{\textrm{sparse}}$ are distributed to entries in the QUBO corresponding to the order of interaction: $Q^{(1)}$ to red, $Q^{(2)}$ to green, $Q^{(3)}$ to blue, $Q^{(4)}$ to yellow. The coefficients of the constraint equations contributes to entries in red text. }
\label{fig:egqubo}
\end{figure*}

\end{document}